\documentclass[11pt]{article}
\usepackage[cp1251]{inputenc}
\usepackage[english]{babel}
\usepackage{amssymb,amsmath}
\usepackage{graphicx}
\usepackage{bm}
\begin{document}

\begin{center}
{\bf Quantum states, symmetry and dynamics in degenerate spin s=1 magnets}
\bigskip

{\bf M.Y. Kovalevsky, A.V. Glushchenko}
\medskip

{\small {\sf Kharkov Institute of Physics and Technology,   \\
     Academicheskaya 1, Kharkov, 61108, Ukraine,\\
     e-mail: $\underline{\mbox{mikov51@mail.ru}}$}}
\end{center}

\begin{abstract}
The article deals with spin s=1 magnets. The symmetry conditions for normal and degenerate equilibrium states are defined and types of magnetic ordering found out. For each type of symmetry breaking the structure of source in the Gibbs statistical operator has been obtained and additional thermodynamic parameters introduced. The algebra of Poisson bracket for magnetic degrees of freedom has been established and nonlinear dynamic equations have been derived. Using the models of the exchange interaction, we have calculated the spectra of collective excitations for two degenerate states whose order parameters have different signature under the time reversal transformation.
\end{abstract}

keywords: quantum states, spin 1, symmetry, dynamics, spectra.

\section{Introduction}
Currently, there is an increasing interest in studies of high spin magnets, which have the spin s$\geq$1. These studies are relevant because of theoretical and experimental work on the physics of quasi-crystalline structures created on the basis of technology of optical lattices \cite{bib1,bib2}. The capability to control geometrical parameters of the lattice and the intensity of the inter particle interaction makes them attractive when studying collective properties of quantum objects. The additional stimulus is associated with the Bose-Einstein condensation of neutral atoms with a non-zero spin \cite{bib3, bib4}. The available data on quadrupole magnetic states \cite{bib5,bib6, bib7} reveal the finiteness of the applicability of traditional physical concepts of magnetism to high-spin systems.
In the papers \cite{bib8,bib9, bib10, bib11} it was investigated the equilibrium states of the spin s=1 magnets and considered models of Hamiltonian with a strong biquadratic interaction. Based on this Hamiltonian, phase states of low-dimensional magnets have been analyzed and the possibility of nematic magnetic states has been predicted. Non-equilibrium processes in normal states of the spin s=1 magnets have been analyzed in \cite{bib12,bib13, bib14}. The authors of these papers used a set of dynamic values corresponding to pure quantum states. In \cite{bib15,bib16}, dynamic equations for magnetic values characterizing mixed states have been obtained. In papers \cite{bib17,bib18} relaxation processes of normal states of magnets with the spin s=1 have been studied. In \cite{bib17} the structure of dissipative fluxes in the dynamic equations has been established for the case of the SU(3) symmetry of Hamiltonian. In \cite{bib18} the nature of collective excitations damping has been found and the importance of the effect of the magnetic symmetry on the relaxation mechanism has been noted.
The description of degenerate magnetic states leads to the expansion of magnetic degrees of freedom. The uniaxial spontaneous symmetry breaking corresponds to the antiferromagnetic case and the magnetic phase of superfluid He$^3$-A, and the biaxial symmetry breaking is observed in spin glasses, the superfluid phase of He$^3$-B [19].

The expected new physical phenomena in spin 1 magnets are mainly due to three factors. With the increase of the particle spin, the set of values required for a macroscopically complete description of ordered magnets states is expanded. The diversity of symmetry properties of $\zeta\varepsilon\eta$ the magnetic                          exchange interaction with s$\geq$1 leads to a more complicated structure of the equilibrium states and non-equilibrium dynamic processes. These magnets have several types of symmetry breaking of equilibrium states due to the different properties of the order parameters under time reversal transformation.

The structure of the paper is as follows: in section 2, using the concept of quasiaverages, we have discussed the properties of equilibrium states with spontaneously broken symmetry and introduced additional thermodynamic parameters. In section 3, we have found subalgebras of the Poisson brackets characterizing normal and degenerate states. In section 4, it has been obtained dynamic equations for two types of degenerate states of magnets and calculated spectra of collective excitations.
\section{Symmetry of normal and degenerate equilibrium states}\label{sec3}
Let us consider mixed quantum equilibrium states of the magnets, the particles of which have spin s=1, and formulate their symmetry properties. In the investigated media, normal equilibrium states have the SO(3) or SU(3) symmetry and are described by the Gibbs statistical operator
\begin{equation}\label{eq8}
\hat{\hat{w}}=exp(\Omega-Y_{a}\hat{\hat{\gamma}}_{a}).
\end{equation}
We have denoted the second quantization operators by   $\hat{\hat{A}}$  to distinguish them from finite-dimensional matrices. In case of the SO(3) symmetry, the exchange Hamiltonian  $\hat{\hat{H}}$ and the spin operator $\hat{\hat{S}}_{\alpha}\equiv-i\varepsilon_{\alpha\beta\gamma}\int{d^3x\hat{\hat{\psi}}_{\beta}^+(x)\hat{\hat{\psi}}_{\gamma}(x)}$  are integrals of motion  $\hat{\hat{\gamma}}_{a}=\hat{\hat{H}}$, $\hat{\hat{S}}_{a}$, ($a=0,\alpha$), acting in the Hilbert space. Here $\hat{\hat{\psi}}_{\beta}^+$, $\hat{\hat{\psi}}_{\gamma}$  are field creation and annihilation operators of particles with the spin s=1. Thermodynamic forces   $Y_{\alpha}$ conjugate of integrals of motion are  $Y^{-1}_0\equiv T$ – temperature and $-Y_{\alpha}/Y_0\equiv h_{\alpha}$ – effective magnetic field. The thermodynamic potential  $\Omega$ is determined from the normalization condition of the Gibbs statistical operator Sp$\hat{\hat{w}}=1$. The operation of taking the trace in the Hilbert space is denoted by Sp to distinguish it from the similar one used for finite-dimensional matrices. The Hamiltonian and the normal equilibrium state satisfy the symmetry conditions \cite{bib20}
\begin{equation}\label{eq9}
\left[\hat{\hat{H}}, \hat{\hat{S}}_{\alpha}\right]=0,~~~ \left[\hat{\hat{w}}, \hat{\hat{\Sigma}}_{\alpha}({\bf Y})\right]=0.
\end{equation}
The operator of the generalized spin moment   $\hat{\hat{\Sigma}}_{\alpha}({\bf Y})$ is given by \\
\begin{center}
$\hat{\hat{\Sigma}}_{\alpha}({\bf Y})\equiv\hat{\hat{S}}_{\alpha}+S_{\alpha}^{\bf Y}$,~ $S_{\alpha}^{\bf Y}\equiv-i\varepsilon_{\alpha\beta\gamma}Y_{\beta}\frac{\partial}{\partial Y_{\gamma}}$
\end{center}
and it acts in both the Hilbert space and space of the thermodynamic force ${\bf Y}$. It satisfies the commutation relations
\begin{equation}\label{eq10}
i\left[\hat{\hat{\Sigma}}_{\alpha}({\bf Y}), \hat{\hat{\Sigma}}_{\beta}({\bf Y})\right]=-\varepsilon_{\alpha\beta\gamma}\hat{\hat{\Sigma}}_{\gamma}({\bf Y}), ~ i\left[s_{\alpha}^{\bf Y}, Y_{\beta}\right]=\varepsilon_{\alpha\beta\gamma}Y_{\gamma}.
\end{equation}
The coincidence of the properties of the SO(3) symmetry of the Hamiltonian and the normal equilibrium state should be understood in terms of relation (\ref{eq9}). Formulas (\ref{eq9}), (\ref{eq10}) show that the equilibrium state is invariant under unitary transformations of the spin rotation  $\hat{\hat{U}}({\bf {\bm \theta}, Y})=exp(i\theta_{\alpha}\hat{\hat{\Sigma}}_{\alpha}({\bf Y}))$, $\hat{\hat{U}}({\bm \theta},{\bf Y})\hat{\hat{w}}\hat{\hat{U}}^+({\bm \theta},{\bf Y})=\hat{\hat{w}}$, where ${\bm \theta}$ is the transformation parameter. In this state, the spin  $s_{\alpha}\left({\bf Y}\right)$ = Sp$\hat{\hat{w}}({\bf Y})\hat{\hat{s}}_{\alpha}$ is collinear to the vector  ${\bf Y}$ and its value in terms of the thermodynamic potential density  $\omega({\bf Y})=\lim_{V\to\infty}\Omega/V$ is given by $s_{\alpha}({\bf Y})=2Y_{\alpha}\partial\omega({\bf Y})/\partial Y^2$. The spin density in the equilibrium state tends to zero $s_{\alpha}({\bf Y})\rightarrow 0$ at ${\bf Y}\rightarrow 0$. This case is similar to the SO(3) symmetric paramagnetic state of spin s=1/2 magnets.

The Gibbs statistical operator of normal equilibrium states of magnets with SU(3) symmetries is also given by (\ref{eq8}). The set of additive integrals of motion $\hat{\hat{\gamma}}_{a}\equiv\left(\hat{\hat{H}}, \hat{\hat{G}}_{\alpha\beta}\right), (a=0, \alpha\beta)$ contains a non-Hermitian matrix operator \cite{bib21}.
\begin{equation}\label{eq11}
\hat{\hat{G}}_{\alpha\beta}=\int{d^3x\left(\hat{\hat{\psi}}_{\alpha}^+({\bf x})\hat{\hat{\psi}}_{\beta}({\bf x})-\delta_{\alpha\beta}\hat{\hat{\psi}}_{\gamma}^+({\bf x})\hat{\hat{\psi}}_{\gamma}({\bf x})/3\right)}
\equiv\hat{\hat{Q}}_{\alpha\beta}+\frac{i}{2}\varepsilon_{\alpha\beta\gamma}\hat{\hat{S}}_{\gamma}.
\end{equation}
Its symmetric part is the quadrupole operator, and the antisymmetric part expressed in terms of the spin operator. Due to the definition (\ref{eq11}) and the commutation relations of second quantization Bose operators, the following relation are true
\begin{equation}\label{eq12}
\left[\hat{\hat{G}}_{\alpha\beta}\hat{\hat{G}}_{\mu\nu}\right]=\hat{\hat{G}}_{\alpha\nu}\delta_{\beta\mu}-\hat{\hat{G}}_{\mu\beta}\delta_{\alpha\nu}.
\end{equation}

Properties of the SU(3) symmetry of the Hamiltonian and the equilibrium state are similar to the formulas (\ref{eq9}):
\begin{equation}\label{eq13}
\left[\hat{\hat{H}}, \hat{\hat{G}}_{\alpha\beta}\right]=0, ~~~ \left[\hat{\hat{w}},\hat{\hat{G}}_{\alpha\beta}\left(\hat{Y}\right)\right]=0,
\end{equation}
where we introduced the operator
\begin{equation}\label{eq14}
\hat{\hat{G}}_{\alpha\beta}\left(\hat{Y}\right)=\hat{\hat{G}}_{\alpha\beta}+G_{\alpha\beta}^{\hat{Y}}, ~~ G_{\alpha\beta}^{\hat{Y}}\equiv Y_{\alpha\lambda}\frac{\partial}{\partial Y_{\beta\lambda}}-Y_{\lambda\beta}\frac{\partial}{\partial Y_{\lambda\alpha}}.
\end{equation}
Thermodynamic forces $Y_{\alpha\beta}$ are conjugate of additive integrals of motion $\hat{\hat{G}}_{\alpha\beta}$. For the operators (\ref{eq14}) the following relations are true
\begin{equation}\label{eq15}
\left[\hat{\hat{G}}_{\alpha\beta}\left(\hat{Y}\right), \hat{\hat{G}}_{\mu\nu}\left(\hat{Y}\right)\right]=\hat{\hat{G}}_{\alpha\nu}\left(\hat{Y}\right)\delta_{\beta\mu}-\hat{\hat{G}}_{\mu\beta}\left(\hat{Y}\right)\delta_{\alpha\nu},~ \left[G_{\alpha\beta}^{\hat{Y}},Y_{\mu\nu}\right]=Y_{\alpha\nu}\delta_{\beta\mu}-Y_{\mu\beta}\delta_{\alpha\nu}.
\end{equation}
Using the formulas (\ref{eq14}) and (\ref{eq15}), it is easy to see that the equilibrium state is invariant under unitary transformation $\hat{\hat{U}}\left(\hat{\theta}, \hat{Y}\right)=\exp i\theta_{\alpha\beta}\hat{\hat{G}}_{\beta\alpha}\left(\hat{Y}\right): \hat{\hat{U}}\left(\hat{\theta}, \hat{Y}\right)\hat{\hat{w}}\hat{\hat{U}}^+\left(\hat{\theta}, \hat{Y}\right)=\hat{\hat{w}}$. To meet hermiticity condition of statistical operator and unitarity condition for the operator $\hat{\hat{U}}\left(\hat{\theta}, \hat{Y}\right)$ the thermodynamic parameters satisfy relations $Y^{\ast}_{\alpha\beta}=Y_{\beta\alpha}, \theta^{\ast}_{\alpha\beta}=\theta_{\beta\alpha}$. Let us present the equilibrium value of the matrix $g_{\alpha\beta} = $Sp$\hat{\hat{w}}\hat{\hat{g}}_{\alpha\beta}$ in terms of the thermodynamic potential: $g_{\alpha\beta}=\partial\omega/\partial Y_{\beta\alpha}$. If the matrix tends to zero $g_{\alpha\beta}\rightarrow0$ at $Y_{\alpha\beta}\rightarrow0$, then this case corresponds to the SU(3) symmetric paramagnetic state of the matter.

The symmetry conditions (\ref{eq9}) and (\ref{eq13}) of the Gibbs statistical operator at degenerate equilibrium states are not true. Degeneracy of the state leads to an additional dependence of this operator from the parameters of the unitary transformation $\hat{\hat{U}}\left({\bm \theta}, {\bf Y}\right)$, or $\hat{\hat{U}}\left(\hat{\theta}, \hat{Y}\right)$. We study such equilibrium states using the concept of quasiaverages \cite{bib22}. In accordance with it, let us define the equilibrium statistical operator by relation
\begin{equation}\label{eq16}
\hat{\hat{w}}_{\nu}\equiv\exp\left(\Omega_{\nu}-Y_{a}\hat{\hat{\gamma}}_{a}-\nu Y_0\hat{\hat{F}}\right).
\end{equation}
The source $\hat{\hat{F}}\equiv\int{d^3x\left(f_{a}({\bf x})\hat{\hat{\triangle}}_{a}({\bf x})+h.c.\right)}$, breaking the symmetry of an equilibrium state, is the linear functional of the order parameter operator $\hat{\hat{\triangle}}_{a}(\bf x)$. Here $f_{a}(\bf x)$ is a function conjugate of the order parameter operator, which defines its equilibrium value in terms of quasiaverages $\triangle_{a}({\bf x})=\left\langle\hat{\hat{\triangle}}_{a}({\bf x})\right\rangle\equiv\lim_{v\to 0}\lim_{V\to\infty}$Sp$\hat{\hat{w}}_{\nu}\hat{\hat{\triangle}}_{a}({\bf x})$. The quasiaverages depend on structure the source $\hat{\hat{F}}$ function. Assignment of a source structure allows describe magnets with different natures of symmetry breaking of equilibrium states.
Let us note that states with the spontaneously broken symmetry are possible in magnets, where there are several magnetic sublattices. In this case, operators of magnetic values will acquire an index of the magnetic sublattice (n): $\hat{\hat{G}}_{\alpha\beta}\rightarrow\hat{\hat{G}}_{\alpha\beta}^{(n)}$. In the multisublattice case, instead of commutation relations (\ref{eq12}), we obtain
\begin{equation}\label{eq17}
\left[\hat{\hat{G}}_{\alpha\beta}^{(n)}, \hat{\hat{G}}_{\mu\nu}^{(m)}\right]=\left(\hat{\hat{G}}_{\alpha\nu}^{(n)}\delta_{\beta\mu}-\hat{\hat{G}}_{\mu\beta}^{(n)}\delta_{\alpha\nu}\right)\delta_{mn}.
\end{equation}
The property of the SO(3) symmetry of the Hamiltonian is given by (\ref{eq9}), where the operator $\hat{\hat{S}}_{\alpha}\equiv\sum_{m}\hat{\hat{S}}_{\alpha}^{(m)}$ has the physical significance of the complete spin moment. In the formula (\ref{eq13}), the operator $\hat{\hat{G}}_{\alpha\beta}\equiv\sum_{m}\hat{\hat{G}}_{\alpha\beta}^{(m)}$ shall be understood as the operator $\hat{\hat{G}}_{\alpha\beta}$. Let us introduce the order parameter operators by the relation $\hat{\hat{\triangle}}_{\alpha\beta}({\bf x})\equiv d_{m}\hat{\hat{g}}_{\alpha\beta}^{(m)}({\bf x})$, where $d_m$ are some constants, which do not simultaneously become zero or one. Formula (\ref{eq17}) leads to the relation
\begin{equation}\label{eq18}
\left[\hat{\hat{G}}_{\alpha\beta}, \hat{\hat{\triangle}}_{\mu\nu}({\bf x})\right]=\hat{\hat{\triangle}}_{\alpha\nu}({\bf x})\delta_{\beta\mu}-\hat{\hat{\triangle}}_{\mu\beta}({\bf x})\delta_{\alpha\nu}.
\end{equation}
Given the connection  (\ref{eq18}) we find
\begin{equation}\label{eq19}
i\left[\hat{\hat{S}}_{\lambda}, \hat{\hat{\triangle}}_{\mu\nu}({\bf x})\right]=\hat{\hat{\triangle}}_{\alpha\nu}({\bf x})\varepsilon_{\alpha\mu\lambda}-\hat{\hat{\triangle}}_{\mu\beta}({\bf x})\varepsilon_{\nu\beta\lambda}.
\end{equation}
This implies the following formulas
\begin{equation}\label{eq20}
i\left[\hat{\hat{S}}_{\lambda},\hat{\hat{\triangle}}_{\mu\nu}^{(s)}({\bf x})\right]=\hat{\hat{\triangle}}_{\alpha\nu}^{(s)}({\bf x})\varepsilon_{\alpha\mu\lambda}+\hat{\hat{\triangle}}_{\alpha\mu}^{(s)}({\bf x})\varepsilon_{\alpha\nu\lambda},~~~~ i\left[\hat{\hat{S}}_{\alpha},\hat{\hat{\triangle}}_{\beta}({\bf x})\right]=-\varepsilon_{\alpha\beta\lambda}\hat{\hat{\triangle}}_{\lambda}^{(s)}({\bf x}).
\end{equation}
Here we introduced the symmetric and antisymmetric parts in the order parameter operator $\hat{\hat{\triangle}}_{\alpha\beta}\equiv\hat{\hat{\triangle}}_{\alpha\beta}^{(s)}-i\varepsilon_{\alpha\beta\gamma}\hat{\hat{\triangle}}_{\gamma}/2$.

Let us consider the source breaking the equilibrium state symmetry given as
\begin{equation}\label{eq21}
\hat{\hat{F}}({\bm \theta})=\int{d^3x\left(\xi_{\alpha}\hat{\hat{U}}({\bm \theta})\hat{\hat{\triangle}}_{\alpha}({\bf x})\hat{\hat{U}}^+({\bm \theta})+h.c.\right)}.
\end{equation}
Here $\hat{\hat{U}}({\bm \theta})\equiv\hat{\hat{U}}^+({\bm \theta}, {\bf Y}=0)$. Constant complex vector $\bm \xi=\bm \xi_1+i\bm \xi_2$ settles the Cartesian coordinate system in the spin space. Relations (\ref{eq20}) lead to the transformation law of  the order parameter operator
\begin{equation}\label{eq22}
\hat{\hat{U}}({\bm \theta})\hat{\hat{\triangle}}_{\alpha}({\bf x})\hat{\hat{U}}^+({\bm \theta})=R_{\alpha\beta}({\bm \theta})\hat{\hat{\triangle}}_{\beta}({\bf x}).
\end{equation}
The orthogonal rotation matrix $R_{\alpha\beta}({\bm \theta})$ is associated with the spin rotation parameter ${\bm \theta}$ by the formula $R_{\alpha\beta}({\bm \theta})\equiv(\exp(\varepsilon\theta))_{\alpha\beta}, \varepsilon_{\alpha\beta\gamma}\theta_{\gamma}\equiv(\varepsilon\theta)_{\alpha\beta}$. The source (\ref{eq21}), considering (\ref{eq22}), is given by
\begin{equation}\label{eq23}
\hat{\hat{F}}\left(\hat{R}\right)=\int{d^3x\left(\xi_{\alpha}R_{\alpha\beta}({\bm \theta})\hat{\hat{\triangle}}_{\beta}({\bf x})+h.c.\right)}
\end{equation}
and corresponds to the biaxial nature of the SO(3) symmetry breaking.

Now, let us consider the source (\ref{eq21}) with a real vector $\bm \xi\equiv\bm \xi_1$:
\begin{equation}\label{eq24}
\hat{\hat{F}}({\bf n})=\int{d^3x\xi_{1\alpha}\hat{\hat{U}}({\bm \theta})\hat{\hat{\triangle}}_{1\alpha}({\bf x})\hat{\hat{U}}^+({\bm \theta})} =\int{d^3xn_{\beta}({\bm \theta})\hat{\hat{\triangle}}_{1\beta}({\bf x})}.
\end{equation}
In this case, it has been a violation of SO(3) symmetry. The Gibbs statistical operator $\hat{\hat{w}}({\bf Y, n})$ additionally depends on the spin anisotropy unit vector (antiferromagnetic vector) $n_{\beta}({\bm \theta})\equiv\xi_{1\alpha}R_{\alpha\beta}({\bm \theta})$. The density of the thermodynamic potential of the Gibbs statistical operator (\ref{eq16}) with the source (\ref{eq24}) is a function of two scalar invariants $\omega=\lim_{v\to 0}\lim_{V\to\infty}\Omega_{\nu}/V=\omega({\bf Y,n})=\omega(Y^2, {\bf Yn})$. A general case, when ${\bf n}\neq0$ and ${\bf Y}\neq0$ characterize the ferrimagnetic ordering. The special case ${\bf n}\neq0$, ${\bf Y}=0$ at $\lim_{Y\to 0}\partial\omega/\partial{\bf Yn}\rightarrow0$ corresponds to the antiferromagnetic ordering. If $\lim_{Y\to 0}\partial\omega/\partial{\bf Yn}\neq0$, at ${\bf n}\neq0$, ${\bf Y}=0$ the ferromagnetic ordering \cite{bib23} is realized. Let us consider the source breaking the SO(3) symmetry of an equilibrium state given by
\begin{center}
$\hat{\hat{F}}({\bm \theta})=\int{d^3x\xi_{\alpha\beta}\hat{\hat{U}}({\bm \theta})\hat{\hat{\triangle}}_{\beta\alpha}({\bf x})\hat{\hat{U}}^+({\bm \theta})}$.
\end{center}
The real matrix $\xi_{\alpha\beta}$ is symmetric and traceless: $\xi_{\alpha\beta}=\xi_{\beta\alpha}, \xi_{\alpha\alpha}=0$. It settles the anisotropy of magnetic degrees of freedom in an equilibrium state. The unitary transformation $\hat{\hat{U}}({\bm \theta})$ transforms the order parameter operator $\hat{\hat{U}}({\bm \theta})\hat{\hat{\triangle}}_{\alpha\beta}({\bf x})\hat{\hat{U}}^+({\bm \theta})=R_{\alpha\lambda}({\bm \theta})R_{\beta\gamma}({\bm \theta})\hat{\hat{\triangle}}_{\lambda\gamma}({\bf x})$. Therefore,
\begin{equation}\label{eq25}
\hat{\hat{F}}\left(\hat{m}\right)=\int{d^3x\xi_{\beta\alpha}R_{\alpha\lambda}({\bm \theta})R_{\beta\gamma}({\bm \theta})\hat{\hat{\triangle}}_{\lambda\gamma}({\bf x})}=\int{d^3xm_{\gamma\lambda}\left({\bm \theta}\right)}\hat{\hat{\triangle}}_{\lambda\gamma}({\bf x}).
\end{equation}
Here $m_{\gamma\lambda}({\bm \theta})$ is a real, symmetric and traceless matrix. In this case, the Gibbs equilibrium statistical operator depends on thermodynamic forces and the symmetric matrix: $\hat{\hat{w}}(\hat{Y}, \hat{m})$.

Finally, let us consider the case where a complete spontaneous breaking of the SU(3) symmetry occurs. Because of (\ref{eq18}), the source in the Gibbs operator given by
\begin{equation}\label{eq26}
\hat{\hat{F}}(\hat{a})=\int{d^3x\xi_{\alpha\beta}\hat{\hat{U}}(\hat{\theta})\hat{\hat{\triangle}}_{\beta\alpha}({\bf x})\hat{\hat{U}}^+(\hat{\theta})} =\int{d^3xa_{\alpha\beta}(\hat{\theta})\hat{\hat{\triangle}}_{\beta\alpha}({\bf x})},
\end{equation}
where $\hat{\xi}^+=\hat{\xi}$, $\hat{a}\left(\hat{\theta}\right)\equiv\hat{D}\left(\hat{\theta}\right)\hat{\xi}\hat{D}^{-1}\left(\hat{\theta}\right)$ and $\hat{D}\left(\hat{\theta}\right)\equiv\exp\left(-i\left(\hat{\theta}\right)\right)$. Sources (\ref{eq23})-(\ref{eq26}) characterize various ways of symmetry breaking of an equilibrium state. In the studied magnets, there are two types of normal equilibrium states with the SO(3) and the SU(3) symmetry and four types of degenerate states. One of them has the SU(3) symmetry broken, and other three – SO(3) symmetry broken.

Under time reversal transformation  $\hat{\hat{T}}$, the spin $\hat{\hat{S}}_{\gamma}$ and the order parameter $\hat{\hat{\triangle}}_{\gamma}$ changes sign:  $\hat{\hat{T}}\hat{\hat{S}}_{\alpha}\hat{\hat{T}}^+=-\hat{\hat{S}}_{\alpha}^*$,  $\hat{\hat{T}}\hat{\hat{\triangle}}_{\alpha}\hat{\hat{T}}^+=-\hat{\hat{\triangle}}_{\alpha}^*$. The asterisk "*" denotes complex conjugation. For the Hamiltonian, quadrupole operator and order parameter operator $\hat{\hat{\triangle}}_{\alpha}^{(s)}$ following relations are true:  $\hat{\hat{T}}\hat{\hat{H}}\hat{\hat{T}}^+=\hat{\hat{H}}^*$,  $\hat{\hat{T}}\hat{\hat{Q}}_{\alpha\beta}\hat{\hat{T}}^+=\hat{\hat{Q}}_{\alpha\beta}^*$,  $\hat{\hat{T}}\hat{\hat{\triangle}}_{\alpha\beta}^{(s)}\hat{\hat{T}}^+=\hat{\hat{\triangle}}_{\alpha\beta}^{(s)*}$. Using this relations and taking into account (17)-(19) we can find the transformation law for the Gibbs statistical operators (1),(9). This allows one to find some thermodynamic values in equilibrium states.
\section{Subalgebras of the Poisson brackets of physical quantities and types of magnetic states}\label{sec4}
In accordance with the approach \cite{bib16}, for the construction of Hamiltonian mechanics let us introduce Hermitian $3\times3$ matrices ($\hat{a}=\hat{a}^+, \hat{b}=\hat{b}^+$), which are canonically conjugate variables of spin s=1 magnets. This means that the following Poisson brackets are true
\begin{equation}\label{eq27}
\begin{split}
&\{b_{\alpha\beta}({\bf x}), b_{\mu\nu}({\bf x'})\} =0, \{a_{\alpha\beta}({\bf x}), a_{\mu\nu}({\bf x'})\} =0, \\ &\{b_{\alpha\beta}({\bf x}), a_{\mu\nu}({\bf x'})\} =-\delta_{\alpha\nu}\delta_{\beta\mu}\delta(\bf x-x').
\end{split}
\end{equation}
In terms of these matrices, we introduce the Hermitian matrix
\begin{equation}\label{eq28}
\hat{g}({\bf x})\equiv i\left\lfloor\hat{b}({\bf x}),\hat{a}({\bf x})\right\rfloor,
\end{equation}
which has the physical significance of the density of the SU(3) symmetry generator. Using the formulas (\ref{eq28}) and (\ref{eq27}), we get the Poisson bracket algebra for this variable:
\begin{equation}\label{eq29}
i\left\{g_{\alpha\beta}({\bf x}), g_{\gamma\rho}({\bf x'})\right\}=\left(g_{\gamma\beta}({\bf x})\delta_{\alpha\rho}-g_{\alpha\rho}({\bf x})\delta_{\gamma\beta}\right)\delta(\bf x-x').
\end{equation}
Formulas (\ref{eq27}) and (\ref{eq28}) allow us to obtain the Poisson bracket of matrices $\hat{a}({\bf x})$  and $\hat{g}({\bf x})$
\begin{equation}\label{eq30}
i\left\{a_{\alpha\beta}({\bf x}), g_{\gamma\rho}({\bf x'})\right\}=\left(a_{\gamma\beta}({\bf x})\delta_{\alpha\rho}-a_{\alpha\rho}({\bf x})\delta_{\gamma\beta}\right)\delta(\bf x-x').
\end{equation}
Due to  (\ref{eq30}) $\left\{tr\hat{a}({\bf x}),g_{\gamma\rho}({\bf x'})\right\}=0$, therefore it can be further assumed that $tr\hat{a}=0$, so that the matrix  $\hat{a}$ contains eight independent variables.

Magnetic degrees of freedom of spin s=1  magnets consist of the spin vector $s_{\alpha}({\bf x})$ and the quadrupole matrix $q_{\alpha\beta}({\bf x})$, which are associated with the matrix $g_{\alpha\beta}({\bf x})$ by the relation
\begin{equation}\label{eq4}
g_{\alpha\beta}({\bf x})\equiv q_{\alpha\beta}({\bf x})-i\varepsilon_{\alpha\beta\gamma}s_{\gamma}({\bf x})/2.
\end{equation}
These variables completely characterize the normal states of the studied magnets. The quadrupole matrix $q_{\alpha\beta}$ is real, symmetric and traceless tensor:  $q_{\alpha\beta}=q_{\beta\alpha}$,  $q_{\alpha\alpha}=0$. In addition to these variables, in degenerate case, the state is also characterized by the matrix $\hat{a}$. It is clear that for vector  $s_{\alpha}({\bf x})$, the following Poisson brackets are true due to (\ref{eq4}), (\ref{eq29})
\begin{equation}\label{eq31}
\left\{s_{\alpha}({\bf x}), s_{\beta}({\bf x'})\right\}=\delta({\bf x-x'})\varepsilon_{\alpha\beta\gamma}s_{\gamma}({\bf x}).
\end{equation}
For the variables  $s_{\alpha}({\bf x})$,  $q_{\alpha\beta}({\bf x})$, we get
\begin{equation}\label{eq32}
\begin{split}
&\left\{s_{\alpha}({\bf x}), q_{\beta\gamma}({\bf x'})\right\}=\delta({\bf x-x'})\left(\varepsilon_{\alpha\beta\rho}q_{\rho\gamma}({\bf x})+\varepsilon_{\alpha\gamma\rho}q_{\rho\beta}({\bf x})\right), \\ &\left\{q_{\alpha\beta}({\bf x}), q_{\mu\nu}({\bf x'})\right\}=\delta({\bf x-x'})s_{\gamma}({\bf x})(\varepsilon_{\gamma\alpha\nu}\delta_{\beta\mu} +\varepsilon_{\gamma\beta\mu}\delta_{\alpha\nu}+ \varepsilon_{\gamma\beta\nu}\delta_{\alpha\mu}+\varepsilon_{\gamma\alpha\mu}\delta_{\beta\nu})/4.
\end{split}
\end{equation}
By a similar way, let us connect the Hermitian matrix  $\hat{a}$ with physical values
\begin{center}
$a_{\alpha\beta}({\bf x})\equiv m_{\alpha\beta}({\bf x})-i\varepsilon_{\alpha\beta\gamma}n_{\gamma}({\bf x})/2.$
\end{center}
Vector $\bf n$ has a physical significance of an antiferromagnetic vector. The tensor $\hat{w}$ has the significance of a T-even order parameter of the nematic ordering. Because of (\ref{eq30}), we obtain the following Poisson brackets
\begin{equation}\label{eq33}
\begin{split}
&\left\{s_{\alpha}({\bf x}), n_{\beta}({\bf x'})\right\}=\delta({\bf x-x'})\varepsilon_{\alpha\beta\gamma}n_{\gamma}({\bf x}), \\ &\left\{n_{\alpha}({\bf x}), q_{\beta\gamma}({\bf x'})\right\}=\delta({\bf x-x'})\left(\varepsilon_{\alpha\beta\rho}m_{\rho\gamma}({\bf x})+\varepsilon_{\alpha\gamma\rho}m_{\rho\beta}({\bf x})\right), \\ &\left\{s_{\alpha}({\bf x}), m_{\beta\gamma}({\bf x'})\right\}=\delta({\bf x-x'})\left(\varepsilon_{\alpha\gamma\rho}m_{\beta\rho}({\bf x})+\varepsilon_{\alpha\beta\rho}m_{\gamma\rho}({\bf x})\right), \\ &\left\{m_{\alpha\beta}({\bf x}), q_{\mu\nu}({\bf x'})\right\}=\delta({\bf x-x'})n_{\gamma}({\bf x})(\varepsilon_{\alpha\nu\gamma}\delta_{\beta\mu}+\varepsilon_{\beta\mu\gamma}\delta_{\alpha\nu}+ \varepsilon_{\beta\nu\gamma}\delta_{\alpha\mu}+\varepsilon_{\alpha\mu\gamma}\delta_{\beta\nu})/4.
\end{split}
\end{equation}
Formulas (\ref{eq31})-(\ref{eq33}) reveal the subalgebras of the Poisson brackets and allow us to determine the dynamics of magnets with the spin s=1 for all the types of ordering. Let us characterize each of them in detail:

Case 1: The minimal subalgebra of the Poisson brackets (\ref{eq31}) contains only the spin vector. The Hamiltonian formalism leads to Landau-Lifshitz equation [24] describing s=1/2 magnets.

Case 2: The magnetic degrees of freedom consist of the spin density and the quadrupole matrix. The dynamic equations of this kind of states have been obtained and analyzed in papers \cite{bib15,bib16}.

Case 3: The set of magnetic dynamic values consists of the spin density and the vector of spin anisotropy, for which a closed subalgebra of the Poisson brackets (\ref{eq31}),(\ref{eq33}) is valid. This case describes uniaxial T-odd SO(3) symmetry breaking with respect to rotations in the spin space. The dynamics of such magnets is equivalent to the antiferromagnet.

Case 4: The magnetic degrees of freedom consist of the spin density and the orthogonal matrix of rotation $\hat{R}$. To determine the Poisson brackets of the last variable with the spin density, let us note that the arbitrary orthogonal matrix can be expressed in terms of a real anti-symmetric matrix $\hat{R}\equiv(1+\hat{\eta})(1-\hat{\eta})^{-1}$. Let us define the matrix $\eta_{\alpha\beta}$ in terms of the matrix $a_{\alpha\beta}$ by relation $\eta_{\alpha\beta}\equiv i(a_{\alpha\beta}-a_{\beta\alpha})$.  By further using the formula (\ref{eq30}), we get
\begin{equation}\label{eq34}
\left\{R_{\alpha\beta}({\bf x'}), s_{\lambda}({\bf x})\right\}=\delta({\bf x-x'})\left(\varepsilon_{\lambda\gamma\alpha}R_{\gamma\beta}({\bf x})-\varepsilon_{\lambda\beta\gamma}R_{\alpha\gamma}({\bf x})\right).
\end{equation}
Formulas (25),(28) are the set of Poisson brackets for the case of  biaxial T-odd breaking of SO(3) symmetry.

Case 5: The spin vector and the tensor $m_{\alpha\beta}({\bf x})$ form a subalgebra of the Poisson brackets (\ref{eq31}),(\ref{eq33}). It is a physically new case of T-even breaking of the SO(3) symmetry, which is absent in spin s=1/2 magnets.

Case 6: A set of magnetic values consists of Hermitian matrices  $\hat{a}$ and $\hat{g}$. Formulas (\ref{eq29}),(\ref{eq30}) allow us to obtain dynamics equations of spin s=1 magnets in condition of complete breaking of SU(3) symmetry. However, in view of inconvenience, we do not consider them in this paper.
\section{Dynamic equations and excitation spectra of degenerate states}\label{sec5}
The main interaction in magnets has an exchange nature. The consideration of dynamic processes requires the formulation of conservation laws in the differential form, taking into account the Hamiltonian symmetry. The condition of the SO(3) symmetry of the exchange energy  density is given by
\begin{equation}\label{eq35}
\left\{S_{\alpha}, e({\bf x})\right\}=0.
\end{equation}
The exchange energy density has the form  $e=e_{hom}+e_{inhom}$. Here the homogeneous part of the energy density depends on the spin density and the variables associated with the broken symmetry. For simplicity, we consider the contribution to the inhomogeneous part of the energy only in the form of gradients matrices  $\hat{R}$ or $\hat{m}$.

Case 4. The Poisson bracket (\ref{eq31}),(\ref{eq34}) and the symmetry condition (\ref{eq35}) lead to the following dynamics equations
\begin{equation}\label{eq36}
\begin{split}
&\dot{s}_{\alpha}=-\nabla_{k}\varepsilon_{\alpha\beta\gamma}\left(\frac{\partial e}{\partial\nabla_{k}R_{\beta\lambda}}R_{\gamma\lambda}+\frac{\partial e}{\partial\nabla_{k}R_{\lambda\beta}}R_{\lambda\gamma}\right), \\ &\dot{R}_{\alpha\beta}=\left(\varepsilon_{\rho\beta\gamma}R_{\alpha\rho}+\varepsilon_{\alpha\gamma\rho}R_{\rho\beta}\right)\frac{\delta H}{\delta s_{\gamma}}.
\end{split}
\end{equation}
We construct the model expression of the exchange energy density for spins s=1 magnets from the Casimir invariant of the Poisson bracket (\ref{eq31}) and Casimir invariants for an expanded set of Poisson brackets (\ref{eq31}),(\ref{eq34}). They are $R_1\equiv tr\hat{R},R\equiv tr\hat{R}^{2}=4cos^2\theta-1, R_3\equiv tr\hat{R}^3$. Since these invariants are connected, we choose  $R$ as a single independent variable. We shall form the exchange energy model so that its homogeneous part had a specific sign, and the inhomogeneous part is a positive. We choose the energy density as follows [25]:
\begin{equation}\label{eq37}
\begin{split}
 &e_{hom}=-\frac{1}{2}As^2-\frac{1}{2}BR^2+\frac{1}{4}Es^4+\frac{1}{4}FR^4++\frac{1}{2}Js^2R^2, \\ &e_{inhom}=\frac{1}{2}D(\nabla_{k}R_{\alpha\beta})^2+\frac{1}{2}C\nabla_{k}s^2.
\end{split}
\end{equation}
Here $A,B,E,F,J$  are effective exchange integrals of the homogeneous magnetic interaction and $D$ is the exchange integral of inhomogeneous interaction. The stability of the equilibrium state in case C=0 1) $s_0=0, R_0=0$ is provided by inequalities $A<0, B<0$. The Goldstone wave spectrum is linear $\omega=k\sqrt{-2AD}|\sin\theta_0|$. 2) The state $ s_0=0, R^2_0=B/F$ is stable, if $F>0, B>0, BN>FA$. The Goldstone wave spectrum is linear:  $\omega=k\sqrt{2D(NB/F-A)}|\sin\theta_0|$; the wave propagates transversely with respect to the axis ${\bm \theta}_0/|{\bm \theta}_0|$. 3) The ferromagnetic state $ s_0^2=A/E, R_0=0$  is stable, if: $A>0, E>0, AN>BE$. The spin wave spectrum is linear  $\omega=2k\sqrt{(s^2_0-({\bf s}_0,{\bm \theta}_0/|{\bm \theta}_0|)^2)DE}|\sin\theta_0|$ and the wave propagates transversely to the direction ${\bf s}_0\times{\bm \theta}_0/|{\bm \theta}_0|$; 4) The equilibrium state $s_0^2=(AF-NB)/(FE-N^2), R^2_0=(ER-AN)/(FE-N^2)$ is stable, if $E>0, AF>BN, EF>N^2, BE>AN$.  The spectrum is linear  $\omega=2k\sqrt{DE(AF-NB)/(FE-N^2)}|\sin\theta_0||\sin\varphi_0|$, where $\varphi_0$ is angle between the spin vector $\bf s$ and the vector $\bm \theta$.

Case 5. Considering formulas (\ref{eq31}),(\ref{eq33}) and the symmetry condition (\ref{eq35}) of the energy density $e=e({\bf s},\hat{m},\nabla\hat{m})$, we get equations
\begin{equation}\label{eq38}
\dot{s}_{\alpha}=-2\nabla_{k}\varepsilon_{\alpha\beta\gamma}\left(\frac{\partial e}{\partial\nabla_{k}m_{\beta\lambda}}m_{\gamma\lambda}\right), ~~~~ \dot{m}_{\beta\gamma}=-\left(\varepsilon_{\alpha\gamma\rho}m_{\rho\beta}+\varepsilon_{\alpha\beta\rho}m_{\rho\gamma}\right)h_{\alpha}.
\end{equation}
Here $h_{\alpha}=\delta H/\delta s_{\alpha}$. The solution of (\ref{eq38}) at the equilibrium point leads to the relations: 1) $h_{\alpha}=0,  \hat{m}$ - const; 2) $h_{\alpha}=hn_{\alpha}$   and  $m_{\alpha\beta}=m(e_{\alpha}e_{\beta}-1/3\delta_{\alpha\beta})$, uniaxial case; 3)  $h_{\alpha}=hl_{\alpha}$,  $m_{\alpha\beta}=m(n_{\alpha}n_{\beta}-f_{\alpha}f_{\beta})$, biaxial case. Vectors $\bf f, n, l=f\times n$ are the orthonormal frame in spin space.

Let us choose the exchange energy density model in the form of (\ref{eq37}) with substitutions $(tr\hat{R}^2)\rightarrow tr\hat{m}^2$ and  $(\nabla R_{\alpha\beta})^2\rightarrow(\nabla m_{\alpha\beta})^2$. It is clear that for this energy model, the following equilibrium states are possible: 1) $s_0=0, m_0=0$ – the paramagnetic equilibrium state is stable, if: $A<0, B<0$; there is no real part of spectrum. 2) The solution $ s^2_0=A/E, m_0=0$  is a stable ferromagnetic equilibrium state, if: $E>0, EB<JA, A>0$. The spin wave spectrum is quadratic  $\omega=Cs_0k^2$; 3) $s_0=0, m^2_0=3B/2F$– the quadrupole equilibrium state (spin nematic) is stable, if: $B>0, F>0, AF<JB$. The quadrupole wave spectrum is given by $\omega=k\sqrt{6DB(-FA+JB+FCk^2)}/F$. 4) Solutions $s^2_0=(AF-BJ)/(EF-J^2), m^2_0=3(BE-AJ)/2(EF-J^2)$,    describe the stable equilibrium state, if $BE>AJ, EF>J^2, AF>BJ, E>0$. The spectrum is linear $\omega=2\sqrt{6DE(BE-AJ)(AF-BJ)}k|\sin\psi|/(EF-J^2)$, where $\psi$ is angle between the spin vector ${\bf s}_0$ and the matrix axis $\hat{m}^0$.

The analysis of the symmetry of the equilibrium magnetic states shows that along with the two types of normal states with SO(3) or SU(3) symmetry, there are three types of degenerate states: two T-odd types of SO(3) symmetry breaking (uniaxial and biaxial vector order parameter) and one T-even state (quadrupole order parameter). To date, not found experimental confirmation of SU(3) symmetry of the equilibrium state in spin 1 magnets. In our work we have shown the possibility of manifestation of the quadrupole degree of freedom in terms of T-even SO(3) symmetry breaking of the equilibrium state of such magnets for which the spectra of magnetic excitations are found.

\bibliographystyle{model1a-num-names}
\bibliography{<your-bib-database>}

\end{document}